\documentclass[aps,prb,twocolumn,showpacs]{revtex4}
\usepackage[dvips]{graphics}
\usepackage{shortvrb}
\MakeShortVerb{\!}

\begin{document}

\title{
Specific   heat  and    high-temperature  series  of  lattice   models:
interpolation scheme  and examples on quantum  spin systems in one and
two dimensions}

\author{B.~Bernu}
\email{bernu@lptl.jussieu.fr}
\affiliation{Laboratoire de Physique
Th{\'e}orique des Liquides, Universit{\'e} P.  et M.  Curie, case 121,
4 Place Jussieu, 75252 Paris Cedex, FRANCE.  URA 765 of CNRS.}

\author{G.~Misguich}
\email{misguich@spht.saclay.cea.fr}
\affiliation{Service de Physique Th{\'e}orique,
CEA Saclay, 91191 Gif-sur-Yvette Cedex, FRANCE}
\bibliographystyle{prsty}

\newcommand{\s}{{\ensuremath{\ln(2)}}}

\begin{abstract}

We have  developed a new  method for  evaluating the specific  heat of
lattice   spin  systems.     It  is   based  on    the  knowledge   of
high-temperature series expansions, the   total entropy of  the system
and the low-temperature expected behavior of the specific heat as well
as the ground-state energy.  By  the choice of an appropriate variable
(entropy   as a function  of  energy),  a stable interpolation  scheme
between  low and high  temperature is performed.  Contrary to previous
methods,  the constraint that the total  entropy is $\log(2S+1)$ for a
spin $S$  on each site   is automatically satisfied.  We  present some
applications  to  quantum spin models  on   one- and  two- dimensional
lattices. Remarkably, in most cases, a good  accuracy is obtained down
to zero temperature.
\end{abstract}
\maketitle
\section{Introduction}

The  accurate knowledge  of  the thermodynamic  quantities  of quantum
magnets is an important issue from the experimental point of view.  It
allows one   to    determine precisely  exchange  energies   $J$  from
experimental  data  or to  identify possible  deviations from  a given
model.  In  this paper we propose a  simple method to compute the heat
capacity from a high-temperature  (HT) expansion.  Numerous techniques
have been developed  to study thermodynamical singularities from power
series expansions.~\cite{guttmann89}  However, in this work, the point
of view is slightly different.  We mainly focus  on the computation of
the  heat capacity  for systems {\em   without a  phase transition} at
$T>0$ (here in   one  and two   dimensions) and  try to  evaluate  the
specific heat $c_v(T)$  accurately in the largest possible temperature
range.    We devised  a  two-point  Pad{\'e}-like  interpolation of  a
particular function, namely the entropy as a function of energy, which
satisfies the energy and entropy sum  rules obeyed by $c_v(T)$.  These
two constraints are non-local in temperature and improve significantly
the convergence of standard Pad{\'e} approximations.

Before   describing the method  itself, let   us  briefly review  some
commonly used methods  to investigate the thermodynamic properties  of
spin systems.

{\bf High-temperature series   expansion}  is the  usual   approach to
evaluate the strength   of microscopic interactions  from experiments.
Thermodynamic quantities such   as magnetic  susceptibility  $\chi$ or
specific heat $c_v$ are expanded  in powers of $\beta=1/T$.  Presently
computers do not allow to compute more than about 20 terms for quantum
systems.\cite{note1}  When analyzed  through  (differential)  Pad{\'e}
approximants, HT series give reliable results for $T$ greater than the
typical   magnetic    exchange energy  of    the    model.  When   the
low-temperature  form of   a   given  quantity  is  known,    Pad{\'e}
approximants can  be  improved  to  reproduce  the  correct asymptotic
behavior.  However, with this approach  it is difficult to investigate
temperatures much below $J$.   In some cases, a detailed understanding
of the low-energy  regime    allows  one to construct    {\em  biased}
approximants and to compute   thermodynamics in a  larger  temperature
range  (see  for instance Ref.~\onlinecite{roger98b}  for the magnetic
susceptibility  of   square-  and   triangular-lattice  ferromagnets).
Another     series    approach     to      thermodynamics   is     the
finite-cluster-expansion.   This  expansion in  powers of the coupling
constant may, in  some cases, allows  one to go to  lower temperature.
It      has  been      applied      in   one~\cite{narayanan90}    and
two~\cite{wang92,zho99} dimensions.

{\bf  Finite-size calculation}.  The  previous approaches are exact at
high  temperature but  do  not satisfy  the sum  rule  that the  total
entropy is $\ln(2S+1)$.  An     alternative is to compute   the   heat
capacity of a finite system from the exact spectrum of a small cluster
of spins, and then compute the partition function  by summing over all
eigenstates (example:   spin-1   chain~\cite{Neef74,nj75}).   A direct
finite-temperature      lancz{\"o}s      algorithm  can     also    be
performed~\cite{jaklic96},  as  well  as  transfer  matrix techniques.
These techniques automatically give  the correct entropy, but they are
limited to  small systems ($N\leq 36$)  and finite-size  errors may be
difficult to control, especially in two dimensions.

Quantitative  extrapolations to $N=\infty$ of finite-size $c_{v,N}(T)$
data have been done in spin chains but are not really efficient at low
temperatures.        It       has      been      applied   to     spin
chains~\cite{BonnerFisher64,nj75,nkk74,blote75}  and the    triangular
antiferromagnet.~\cite{imada87}     Power-law    behaviors   at    low
temperature cannot be observed  due to the important discretization of
the low energy spectrum in a small system.

{\bf Quantum Monte-Carlo} simulations  can reach a rather large number
of  spins    for    not   frustrated     systems  (as    the    square
lattice~\cite{makivic91}).   These  calculations reproduce the correct
HT behavior (up to small  statistical errors) but cannot
reach the  very low temperature regime.  This technique is one  of the
most efficient in  studying thermodynamics properties  but it requires
important numerical  effort and is limited  to small  systems when the
model contains frustration.

{\bf Sum rules}.  We propose a {\em simple} method to compute the heat
capacity which involves a new kind of HT series analysis.  In contrast
to  previous    methods, we  provide  a  procedure  for $c_v(T)$ which
satisfies a first sum rule:
\begin{equation}
	\int_{0}^{\infty}\frac{c_v(T)}{T}dT=\ln(2S+1).
\end{equation}
Moreover,  we incorporate  the  knowledge of  the ground  state energy
$e_0$ and the  energy $e_m$ at  infinite temperature.   These energies
are either known  exactly or can be computed  numerically. We use this
information in a second sum rule:
\begin{equation}
	\int_{0}^{\infty}c_v(T) dT=e(T=\infty)-e(T=0)
\end{equation}

Finally, the low-temperature leading contribution of $c_v(T)$ is taken
into account. 

We   claim that  the   implementation  of these  integral  constraints
increases  in a significant  way  the  range  of validity  of  the  HT
series.\cite{note2}  We will show that this method is indeed successful
even at low or zero  temperature for the  models we have investigated.
By  ``successful'', we  mean that the  specific heat  is obtained {\em
down to zero temperature}  with a relative accuracy typically  between
1\% and 0.1\%  with  only ten  terms in the  high-temperature  series.
This accuracy should    be   considered  as surprisingly  high     for
low-temperature quantities  obtained  with  a high-temperature  series
expansion.

\section{Interpolation procedure}

\subsection{Elementary thermodynamics: From $c_v(T)$ to $s(e)$}

\begin{figure}
\begin{center}
\resizebox{6cm}{!}{\includegraphics{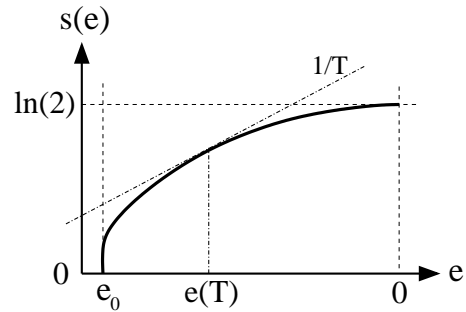}}
\end{center}
\caption{Typical shape of the entropy $s$ as a function of energy $e$
for a spin-$\frac{1}{2}$ system.}
\label{fig:schemaS(E)}
\end{figure}

In  the canonical  ensemble, the energy   per site  $e$  and the  {\em
entropy   per site}   $s$ are   increasing  functions of   temperature
$T=\frac{1}{\beta}$.  Since $e$ is a monotonic function of $T$, we can
express  $s$ as  a function of  $e$  rather than of  temperature.  The
specific  heat as a function  of temperature  is easily expressed with
$s(e)$ in the following way. The entropy $s(T)$ obeys:
\begin{equation}
	\frac{ds}{dT}=\frac{c_v}{T}
\end{equation}
Where $c_v$ is the specific heat per site. It follows that :
\begin{equation}
	s'(e)=\frac{ds}{de}=\frac{c_v}{T}\frac{dT}{de}
\end{equation}
where the  prime denotes differentiation with  respect  to $e$.  Since
$\frac{dT}{de}=\frac{1}{c_v}$,  we have  (the Boltzmann constant $k_B$
is set to unity in the whole paper) :
\begin{equation}
	s'(e)=\frac{1}{T}=\beta		\label{eq:Defbeta}
\end{equation}
We use this relation to eliminate the variable temperature in $c_v$:
\begin{eqnarray}
c_v&	=&\frac{de}{dT}
	=\left[\frac{dT}{de}\right]^{-1}
	=\left[\frac{d}{de}\left(\frac{1}{s'}\right) \right]^{-1}
\end{eqnarray}
We eventually find
\begin{equation}
	c_v=-\frac{s'^2}{s''}		\label{eq:DefCv}
\end{equation}
Eqs.~(\ref{eq:Defbeta},\ref{eq:DefCv}) are the basic relations that we
shall use here. These relations  can also  be obtained by  considering
the density of states of a large but finite system.\cite{note3}.
For simplicity,  we discuss the  case of a  quantum spin-$\frac{1}{2}$
model on a lattice.\cite{note4}  The simplest cases are the Heisenberg
models:
\begin{equation}
	H=2J\sum_{<i,j>} \vec{S}_i\cdot\vec{S}_j
	\label{eq:Hamiltdef}
\end{equation}
where the   sum runs over   the first   nearest neighbors.  $s(e)$  is
defined inside an interval going from the ground-state energy $e_0$ up
to the high-temperature energy $e_m$.   $e_m$ is the free-spin average
of the energy, which is obtained  straightforwardly since, at infinite
temperature one has $\left<\vec{S}_i\cdot\vec{S}_j\right>=0$.  For the
Hamiltonian (\ref{eq:Hamiltdef}), one   simply gets  $e_m=0$.  If  the
ground    state   $\left|0\right>$     is    ferromagnetic ($J<0$   in
Eq.~(\ref{eq:Hamiltdef})),                  one                    has
$\left<0\right|\vec{S}_i\cdot\vec{S}_j\left|0\right>=\frac{1}{4}$  and
$e_0=J\frac{z}{4}$ for  a  Bravais   lattice  of coordination   number
$z$. However, for antiferromagnetic models, $e_0$ is not known exactly
but  Monte-Carlo   simulations,  exact  diagonalizations or analytical
calculation (spin-wave, mean-field Schwinger-bosons, etc) can be used.
In summary  the function $s(e)$, defined in the interval $[e_0,0]$
($e_0<0$), is an increasing function of $e$,  starting at $0$ in $e_0$
with   an infinite slope, and   finishing  at $\ln(2)$  in $e=0$  (see
Fig.~\ref{fig:schemaS(E)}).

To find  an approximation of $s(e)$,  we combine  three types of exact
informations on $s(e)$:
\begin{itemize}
\item[{\bf a)}] The entropy per spin is $\ln(2)$ at infinite temperature
\begin{equation}
	s(e_m=0)=\ln(2)
	\label{eq:s=ln2}
\end{equation}
and vanishes at zero temperature~\cite{note5}:
\begin{equation}
	s(e_0)=0
	\label{eq:s=0}
\end{equation}

\item[{\bf b)}] The behavior of $c_v(T\to \infty)$ is known
from HT series.  From the    expansion of $c_v(T)$  in
powers of  $1/T$ up to order
$1/T^n$ , we get the  expansion  of $s(e)$  in the vicinity of
$e\to e_m=0$ in powers of $e$ to order $e^n$:
\begin{equation}
	s(e)_{e\to0}=	\ln(2) + \sum_{i=2}^n a_i e^i
	\label{eq:sHT}
\end{equation}
This      expansion    can      be       computed       by     solving
Eqs.~(\ref{eq:Defbeta},\ref{eq:DefCv}) order by order for $s(e)$
(details in appendix A).

\item[{\bf c)}] When the low-energy physics of the model is understood,
the low-temperature limit of the specific heat can often be predicted.
In the case where the system exhibits some ferro- or antiferromagnetic
long-range order at  zero  temperature, the low-lying excitations  are
spin waves.  These gapless modes  give a low-temperature heat capacity
which is a power law.  When the  space dimension is $D$, a ferromagnet
has   $c_v(T)\sim T^{D/2}$   and  an  antiferromagnet  has $c_v(T)\sim
T^{D}$. In both  a cases, we can write:
\begin{equation}
\label{eq:cvTpower}
	c_v(T)_{T\to0} \sim T^{\frac{p}{q}}
\end{equation}
where  $p$  and $q$  are   integers.   This low  temperature  behavior
translates into a behavior of $s(e)$ about the ground-state energy per
site $e_0$.
\begin{equation}
	s(e)_{e\to e_0}\sim (e-e_0)^{\frac{p} {p+q}}
	\label{eq:s_e0}
\end{equation}
On the other hand, if elementary excitations are gapped
the system has a thermally activated specific heat
\begin{equation}
\label{eq:cvgap}
	c_v(T)_{T\to0} \sim \exp{\left(-\Delta/T\right)}
\end{equation}
and Eq.~(\ref{eq:s_e0})  is replaced  by a logarithmic  behavior about
the ground-state energy (see appendix~C).

\end{itemize}

\subsection{Interpolation by Pad{\'e} approximants}

We now define the interpolation procedure  between low and high energy
for $s(e)$.  We  look for an approximation  of $s(e)$  which satisfies
Eqs.~(\ref{eq:s=ln2}),      (\ref{eq:s=0}),      (\ref{eq:sHT})    and
(\ref{eq:s_e0}).  A two-point Pad{\'e}  interpolation is  not directly
possible since $s(e)$ is singular  in  $e=e_0$.  We have to  transform
$s$ to obtain regular behavior  in the {\em whole} interval $[e_0,0]$.
This   is  possible when   $p$ and $q$   are two  integers  (but other
low-temperature  form will be  used  for thermally activated $c_v$  in
gapped systems,   see section~\ref{sec:cvgap}).    We  define   a  new
function $G(e)$ as:
\begin{equation}
\label{eq:DefG(e)}
G(e)=s(e)^{p+q}
\end{equation}
This  function now behaves  as  $(e-e_0)^p$ when  $e \to  e_0$  and as
$\s^{p+q}+\mathcal{O}(e^2)$ when $e\to 0$.  It is now possible to look
for approximations $G^{\rm  app}$ of  $G$  which are of  Pad{\'e} form
(details in  appendix~B).  This interpolation scheme   fails if any of
the functions $G(e)$,  $s(e)$, $s^\prime(e)$ or $-s^{\prime\prime}(e)$
becomes negative  in the interval $[e_0,0]$.   This provides a natural
criterion  to     select the  degrees    $[u,d]$  for   which Pad{\'e}
approximants give physical solutions.  The curve for the specific heat
$c_v(T)$  is      eventually   obtained   in    a   parametric    form
$\left\{T(e),c_v(e)\right\}_{e\in[e_0,0]}$           thanks         to
Eqs.~(\ref{eq:DefCv}) and (\ref{eq:DefG(e)}).

\section{Low-temperature power-law behavior of $c_v(T\to0)$}

We start by illustrating the   method for systems where the   specific
heat  is proportional  to $T^\alpha$ at   low temperature.  Some cases
where the  specific heat is thermally activated  will  be described in
the  next section.  Data files  with  the numerical results  presented
here (as  well as colored versions of  the figures) are available upon
request.

\subsection{One-dimensional $S=\frac{1}{2}$ XY model}

As a first application, we consider the spin-$\frac{1}{2}$ XY chain:

\begin{equation}
	H=2\sum_i \left(S^x_i\cdot S^x_{i+1}+S^y_i\cdot S^y_{i+1}\right)
\end{equation}

\begin{figure}
\begin{center}
\resizebox{8cm}{!}{\includegraphics{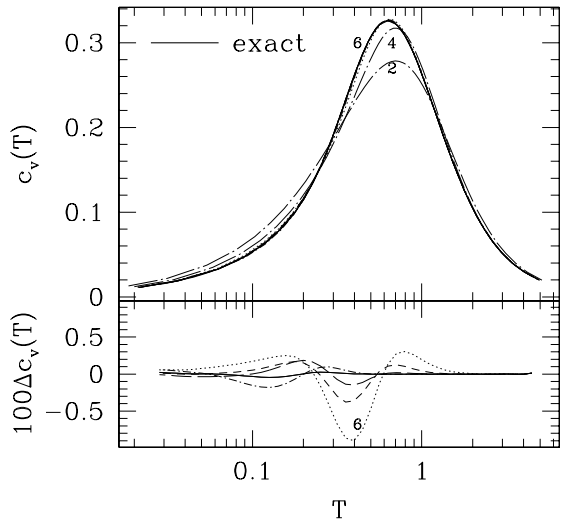}}
\end{center}
\caption{{\bf One-dimensional $S=\frac{1}{2}$ XY model}.
{\em Top}:  Exact specific heat (full  line), approximations at orders
$n=2$ and $n=4$ (dash-dot  lines), $n=6$ (dotted line).  {\em Bottom}:
Differences with the exact result (notice the magnified scale).  $n=6$
(Dotted  line),   $n=8$  (dash  line),  $n=10$   (long  dash),  $n=22$
(dash-dot),   $n=30$ (full  line).   All   Pad\'e  approximants are
diagonal: $[n/2,n/2]$.}
\label{fig:XY1D}
\end{figure}

This spin problem     can   be solved exactly    (the    Jordan-Wigner
transformation  maps   the XY model  to  free  spinless fermions) and
provides a check for a gapless spectrum.  The energy per site at $T=0$
(resp. at $T\to\infty$) is $e_0=-\frac{2}{\pi}$  (resp. $e_m=0$).  The
heat capacity is given by:~\cite{k62}
\begin{equation}
	c_v(\beta)=\frac{2\beta^2}{\pi}
	\int_0^{\pi/2}{\frac{\cos^2(k)}{\cosh^2(\beta\cos(k))}dk}
\end{equation}
From this formula,   the high- and  low-temperature  expansions can be
computed.  We find: $ c_v(\beta\to0)=
\frac{1}{2}\beta^2 -\frac{3}{8}\beta^4 +\cdots$ and
$c_v(T\to0) =\frac{\pi}{6}T +\mathcal{O}(T^3)$.  The linear law at low
temperature          gives        $p=q=1$ in   Eq.~(\ref{eq:DefG(e)}).
Fig.~\ref{fig:XY1D} shows the  comparison  between the  exact $c_v(T)$
and the  ones obtained from  the Pad{\'e} approximants.  At each order
$n$, a few  approximants $[u,n-u]$ lead  to  the same  variations.  In
Fig.~\ref{fig:XY1D},     only   diagonal  Pad{\'e}        approximants
$[\frac{n}{2},\frac{n}{2}]$ are shown.   A convergence of the specific
heat is obtained with a relative accuracy  of the order of one percent
{\em  down  to  zero  temperature} with   only  six  terms in  the  HT
expansion.   We  want     to   emphasize  that  the   slope   $\frac{d
c_v}{dT}(T=0)$ is not imposed in our procedure.  Therefore, the method
provides quantitative information  on this low-energy  parameter which
characterizes the low-energy  excitations.  The value of the prefactor
is exactly  known for this  XY  model and we   find that it oscillates
around the exact value ({\it i.e.} $\pi/6$). It is a few percent below
the exact value at the order $n=6$.

\subsection{
One-dimensional antiferromagnetic Heisenberg spin-$\frac{1}{2}$ model}

\begin{equation}
	H=2\sum_i \vec{S}_i\cdot\vec{S}_{i+1}
\end{equation}

This   model is solvable with the    Bethe ansatz.~\cite{bethe31}  The
ground-state    energy   is     exactly  known~\cite{h38}   to      be
$e_0=-2\ln(2)+\frac{1}{2}$.   The    low-temperature  limit  of    the
heat-capacity  is~\cite{a86}    $c_v(T)=\frac{T}{3}+\mathcal{O}(T^2)$.
The HT expansion   computed  by Baker~{\it et    al.}~\cite{brg64} was
recently extended    to  order  $\beta^{24}$  by   B{\"u}hler~{\it  et
al.}.~\cite{beu00}  The  heat   capacity curve has been   computed for
anisotropic versions of the spin-$\frac{1}{2}$ chain~\cite{takahashi},
but we are   not  aware of  any   exact computation of $c_v$  for  the
isotropic  Heisenberg model based   on  the {\em  exact}  Bethe ansatz
equations (due  to  the existence  of  an infinite  number of  coupled
equations at the isotropic point).   However, an approximate  solution
was proposed  recently by Kl{\"u}mper.~\cite{klumper}  The accuracy of
specific heat obtained by  his method is claimed  to be extremely high
($\sim 10^{-7}$).

\begin{figure}
\begin{center}
\resizebox{8cm}{!}{\includegraphics{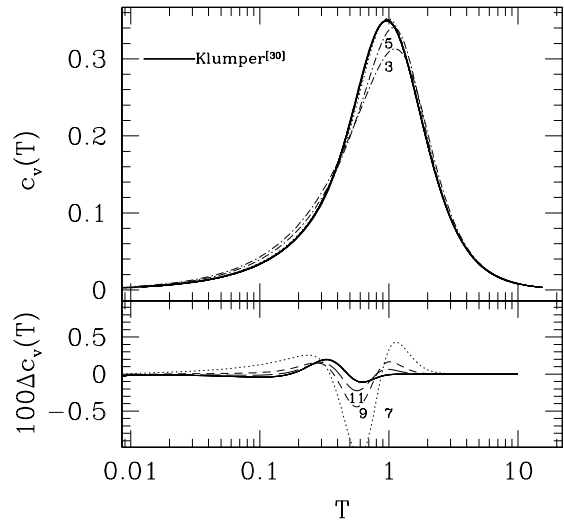}} 
\caption{
Specific  heat of    the   antiferromagnetic  {\bf  spin-$\frac{1}{2}$
Heisenberg  chain}.    {\em Top}:  Result  from  Kl{\"u}mper  {\em et.
al}~\cite{klumper} (full  line), approximations  at orders   $n=3$ and
$n=5$ (dash-dotted lines) and $n=7$ (dots).  {\em Bottom} : Difference
between Ref.~\cite{klumper} and present  approach.  $n=7$ (dot  line),
$n=9$ (dash  line), $n=11$ (long dash),  odd $n$ from $n=13$ to $n=23$
and $n=24$ (full lines).}
\label{fig:Bethe}
\end{center}
\end{figure}

Our results for this model are displayed Fig.~\ref{fig:Bethe}.  A good
convergence is obtained down to  zero temperature.  If we truncate the
HT expansion at order seven, an absolute precision  of one percent can
already be  obtained.  At the highest order,  the specific  heat is in
excellent agreement with the result of Ref.~\onlinecite{klumper}.  The
height  of  the   peak, for  instance,  agrees   with an  accuracy  of
$10^{-4}$.  The low  temperature limit is also  in good agreement with
the exact  result: we find  $c_v(T)\simeq0.329 T$ at order 24 (instead
of  $c_v(T)=\frac{1}{3}T$).      The   difference with   result     of
Ref.~\onlinecite{klumper} at  intermediate temperatures (see bottom of
Fig.~\ref{fig:Bethe}) is   due  to the  fact   our the low-temperature
coefficient  is off by 1\% from  the  exact value $\frac{1}{3}$.  This
result  is obtained without  any  extrapolation when  $n \to  \infty$.
Taking into account finite-order corrections could improve further the
accuracy.

\subsection{Triangular lattice Heisenberg models}

\begin{figure}
\begin{center}
\resizebox{8cm}{!}{\includegraphics{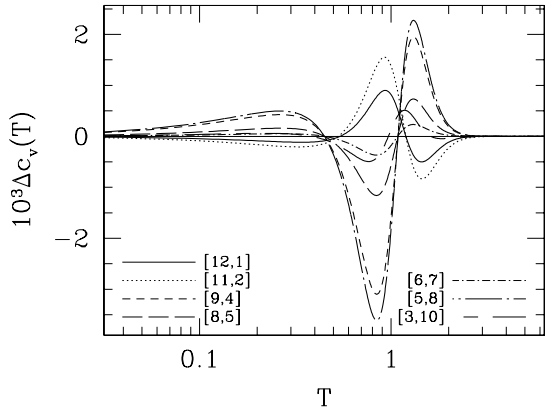}}
\end{center}
\caption{
{\bf   Ferromagnetic   spin-1/2  Heisenberg model   on  the triangular
lattice}. Comparison the specific heats obtained from different Pad\'e
approximants at order $n=13$.  The Pad\'e approximant of reference has
degrees $[7,6]$.  Approximants $[6,7]$, $[8,5]$, $[12,1]$ and $[3,10]$
differ from the reference by less than $10^{-3}$.}
\label{fig:TrFerropadecomp}
\end{figure}

First, we look at the convergence  of this method on the ferromagnetic
case    where    the   ground    state   energy    is   known exactly.
Fig.~\ref{fig:TrFerropadecomp} shows   the  comparison of  the various
Pad{\'e}  interpolants  $[u,d]$,  with  $n=u+d=13$  (highest available
order~\cite{ElstnerHT}).  Two of them have a singularity ($[10,3]$ and
$[4,9]$).  If  we  exclude the polynomial  form  ($d=0$) and the cases
$d>10$, all interpolations lead to  the same variations of $c_v(T)$ in
the  whole range of  temperature.  Fig.~\ref{fig:TrFerropadecomp} also
illustrates the very  small   dispersion  of the different    Pad{\'e}
approximants  obtained with this  method.  It is  also remarkable that
the coefficient of the dominant term at low temperature depends weakly
on the interpolation function used.  As  before, only the power of $T$
is imposed,  but the prefactor is not.   We get $c_v(T)\sim .142(2) T$
at low temperature.  Notice that this quantity  is not given correctly
by a linear spin-wave approximation (non interacting magnons).  We are
not aware of  any previous result  for this  quantity.  A high-density
mono-layer  of  solid $^3$He  solid is,  to   our knowledge,  the only
experimental  realization of  a  triangular-lattice spin-$\frac{1}{2}$
ferromagnet.  The heat  capacity has been  measured by Ishida {\it et.
al}~\cite{ishida97} and their results are  in very good agreement with
our calculations.

\begin{figure*}
\begin{center}
\resizebox{\textwidth}{!}{
\includegraphics{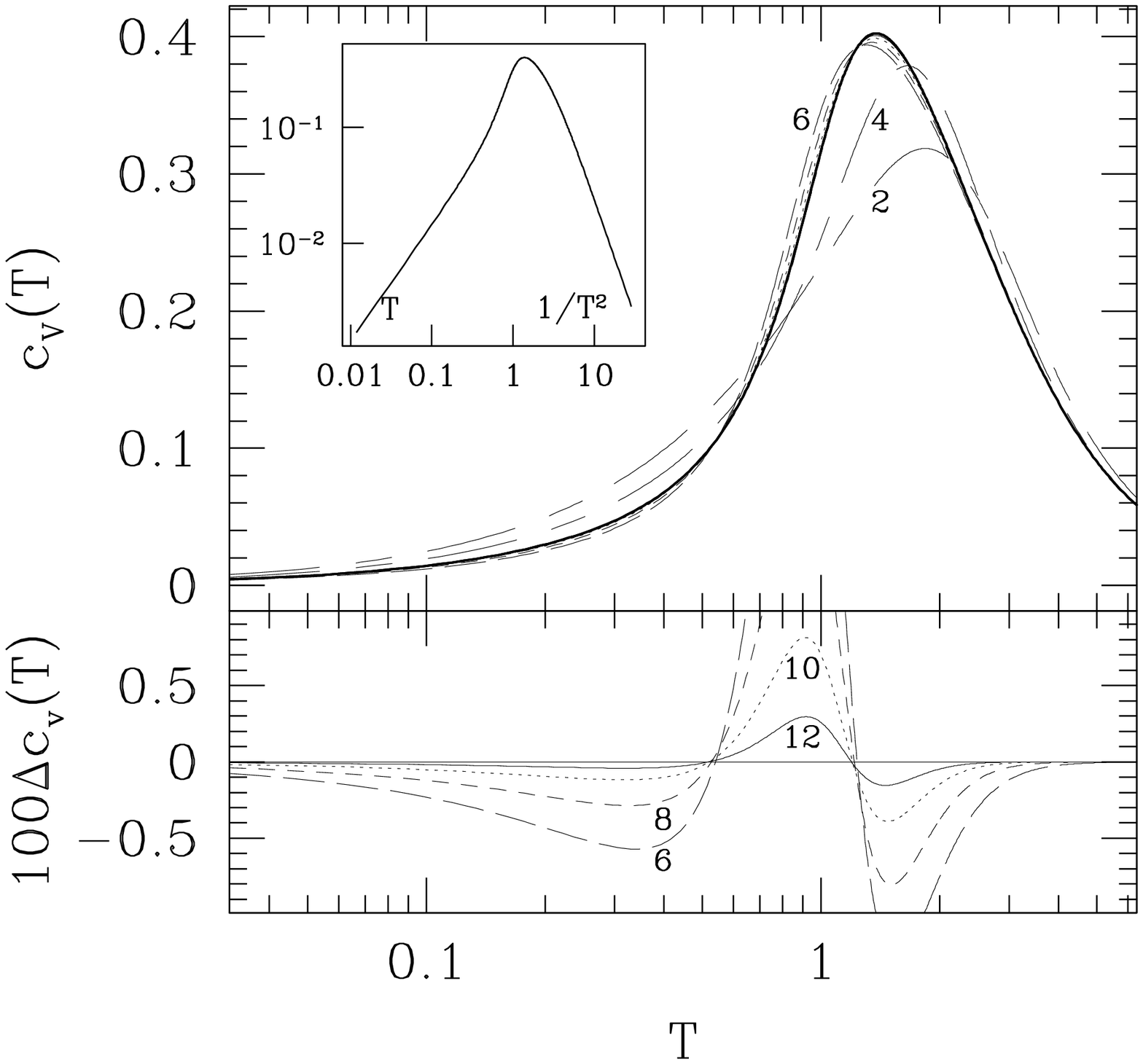}
\includegraphics{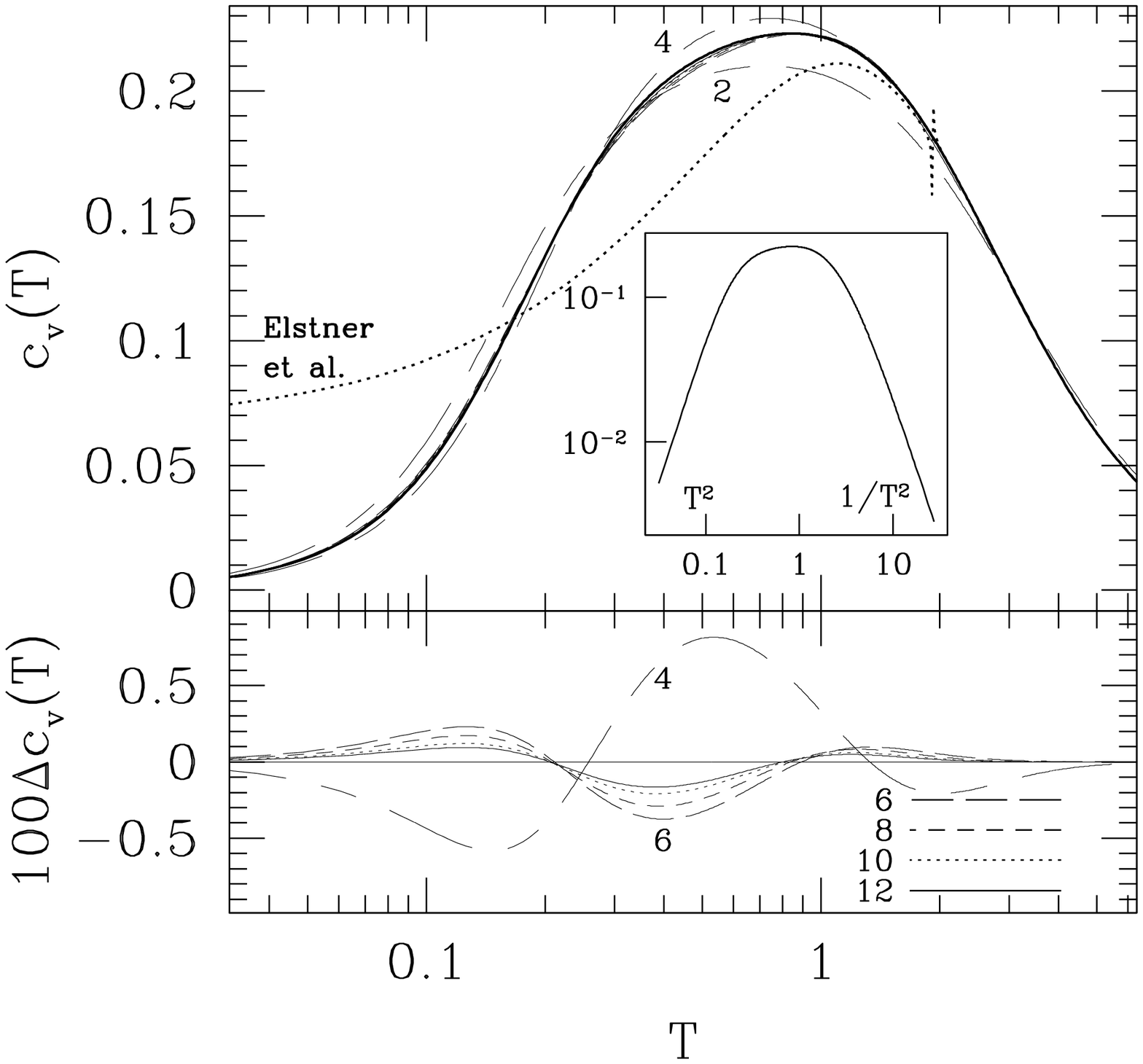}
}
\end{center}
\caption{{\bf Spin-1/2 Heisenberg model on the triangular lattice}. 
{\bf (a)} - {\em   Top}: Variations of  $c_v(T)$  with respect to  the
number of terms in  the HT expansion  in the {\bf ferromagnetic} case.
The even $n$ from 2 to 8 and $n=13$ are shown.  Inset: data for $n=13$
in a  log-log plot.  {\em Bottom}:  Difference  with the highest order
($n=13$). Even $n$ from 6 to 12 are shown.  Order 12  and 13 differ by
less than $3.10^{-3}$.     {\bf (b)}:   Same   as (a)  for  the   {\bf
antiferromagnetic} case.   Order  12   and  13  differ  by  less  than
$2.10^{-3}$.   The dotted line (with the   spike) is from Elstner {\it
el. al}.~\cite{ElstnerHT}}

\label{fig:TrHTcomp}
\end{figure*}

In Fig.~\ref{fig:TrHTcomp}.a and Fig.~\ref{fig:TrHTcomp}.b, we see now
how  the   result converges  when  more  terms are  added    in the HT
expansion.   For each  HT expansion  order  $n$  (as indicated  in the
figure), all possible fractions  $[u,d]$ ($u+d=n$) are tried.   Except
for the lowest order, at each  order several approximants fall on the
``same'' curve.  A reasonable convergence is obtained even for $n=5$.

Fig.~\ref{fig:TrHTcomp}.b   shows   the   approximated   $c_v(T)$ with
increasing HT-order $n$  for  the antiferromagnet Hamiltonian.    This
model    is       frustrated,   but       it     is       now     well
established~\cite{bllp94,lblp95a}     that  the    ground-state  is  a
three-sublattice N{\'e}el state.  The low temperature specific heat is
thus proportional to $T^2$. The ground-state energy was estimated from
exact  diagonalization   data.~\cite{bllp94}  Again   here, we   see a
convergence  for $n>5$.  This should be  compared with direct Pad{\'e}
approximants  to   the     specific    heat,  which    was   done   in
Ref.~\onlinecite{ElstnerHT}           (dotted       curve           in
Fig.~\ref{fig:TrHTcomp}.b).   The comparison   with  our results shows
indeed  that the direct approach does  not even allow to get correctly
the maximum of $c_v(T)$. Our method converges with a relative accuracy
of  about 1\%
down to zero    temperature and  we get   $c_v(T)\sim
5.3(2)T^2$.   It  would  be  interesting to compare   this result with
spin-wave calculations taking magnon-magon interactions into account.

\subsection{Square lattice Heisenberg model}

\begin{figure*}
\begin{center}
\resizebox{\textwidth}{!}{
\includegraphics{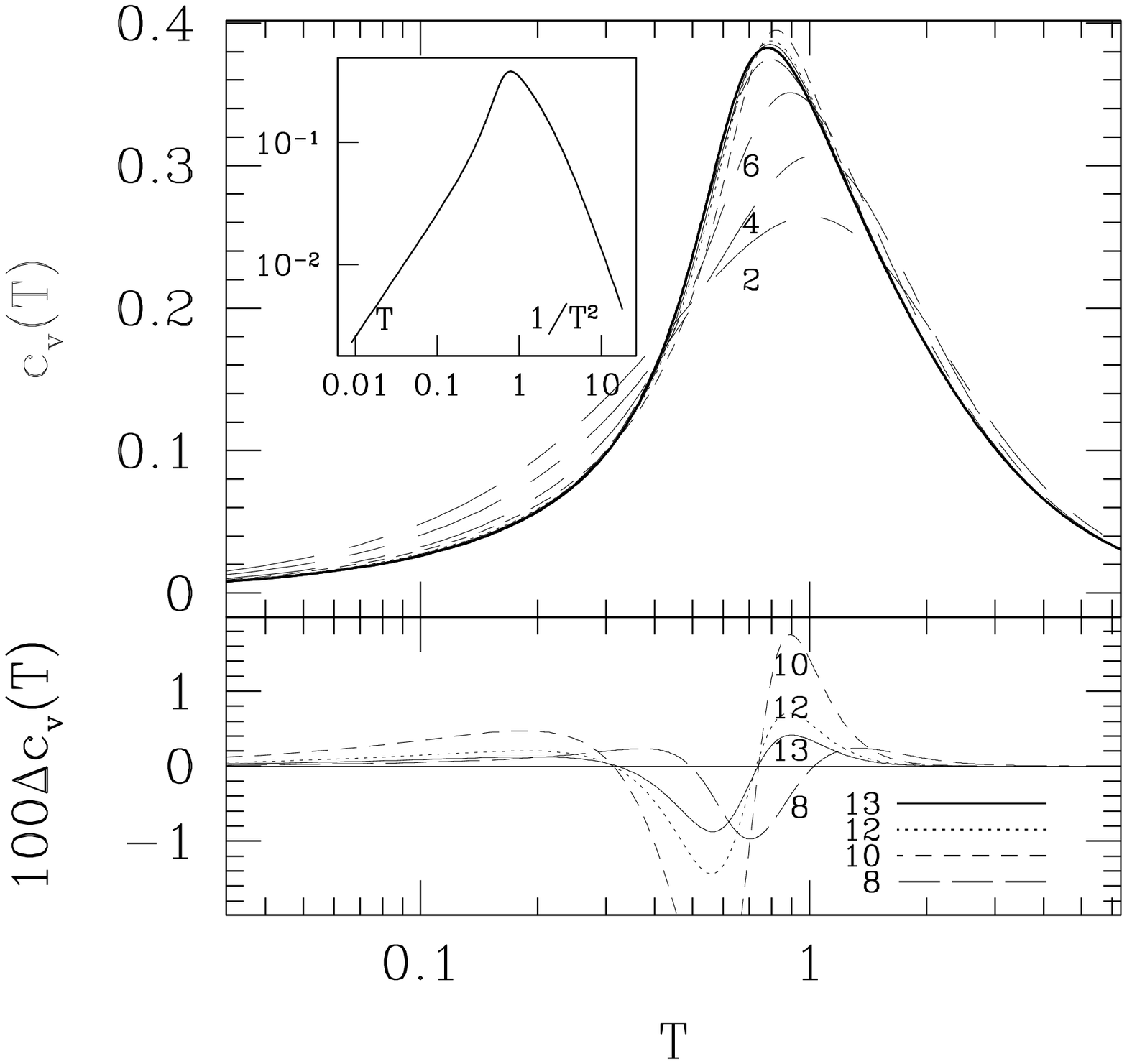} 
\includegraphics{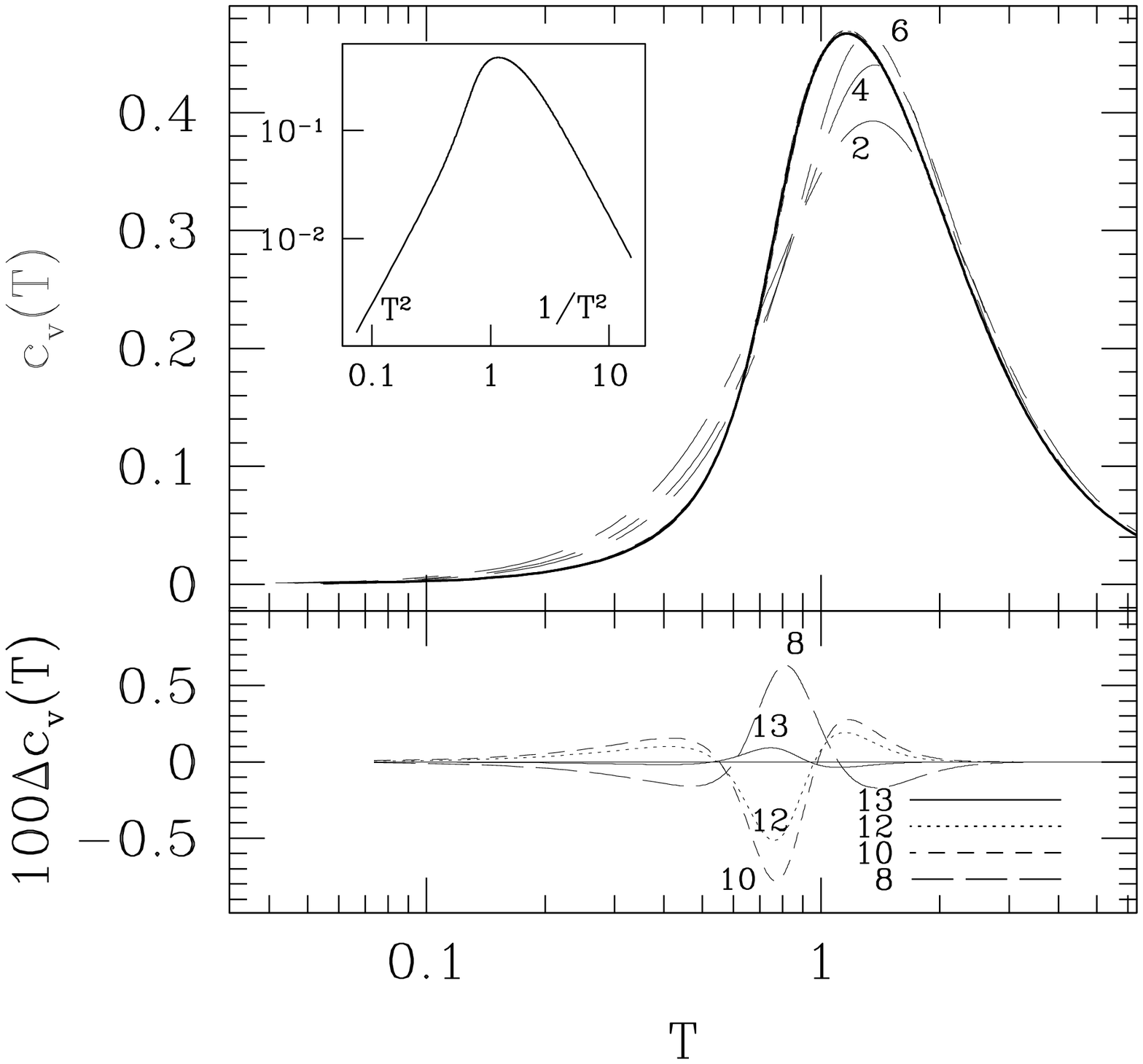}
} 
\caption{spin-$\frac{1}{2}$ Heisenberg model on the {\bf square lattice}.
Same as Fig.~\ref{fig:TrHTcomp} for  the {\bf (a)} square lattice {\bf
ferromagnet} and {\bf (b)} square lattice {\bf antiferromagnet}.  Even
$n$ from 2 to 14 and $n=13$ are shown.  Orders $n=13$ and 14 differ by
less than  $4.10^{-3}$  in the ferromagnetic   case and  by  less than
$10^{-3}$ in the antiferromagnetic case.}
\label{fig:SquareHTcomp}
\end{center}
\end{figure*}

We evaluated the  specific   heat for the  $S=\frac{1}{2}$  Heisenberg
model on  the  square lattice.  Fig.~\ref{fig:SquareHTcomp}  shows the
convergence of  the specific heat with respect  to the number of terms
in the HT-expansion.  The convergence  with the HT-expansion order  is
faster in the  antiferromagnetic case than  in the  ferromagnetic one.
At  low    temperature    we obtain   $c_v(T)=0.25(0.01)    T$  (resp.
$c_v(T)=0.25(0.01)        T^2$)   for   the      ferromagnetic  (resp.
antiferromagnetic) case.  The result in the antiferromagnetic case can
be     compared    with    the     modified     spin-wave   theory  of
Takahashi~\cite{takahashi89} which is $c_v(T)=0.214 T^2$ in our units.

\begin{widetext}
\begin{table*}
\begin{center}
\begin{tabular}{|c|c|c|c||c|c|c|}
\hline
Model 		&F/AF	&$e_0-e_m$		& n			&$c_v(T\to0)$	&$T^{\rm max}$	&$c_v^{\rm max}$\\
\hline
\hline
Heisenberg	&AF	&$-2\ln{2}+\frac{1}{2}$	& 24 \cite{beu00}	&$0.329^*T$	& 0.9618(2)	& 0.3497	\\
chain $S=\frac{1}{2}$&	&\cite{h38}		&			&		&		&		\\
\hline 
Heisenberg	&AF	& -1.401 \cite{wh93}	& 20 \cite{ejg94}	&$\exp{(-\Delta/T)}$& 0.861(1)	& 0.543(1)	\\
chain $S=1$	&	&			&			&$\Delta=.40^*$	&   		&		\\
\hline
Heisenberg	&F	& -3/2			& 13 \cite{ElstnerHT}	&$0.142(2)T$	&  1.375(5)  	& 0.403(3)	\\
\cline{2-3}\cline{5-7}
triangular  lat.&AF	& -1.11 \cite{bllp94}	&			&$5.3(2)T^2$	&  0.84(1)	& 0.2231(5)	\\
\hline
Heisenberg	& F	& -1			& 14 \cite{Oitmaa}	&$0.25(0.01)T$	&  0.785(4)	& 0.383(3)	\\
\cline{2-3}\cline{5-7}
square   lat.	& AF	& -1.34			&			&$0.25(0.01)T^2$& 1.163(2)	& 0.467(2)	\\
		&	&\cite{tc90,bs98}	&			&		&		&		\\
\hline
\end{tabular}
\caption{Parameters of the different models.
Column two indicates whether  the  couplings are ferromagnetic  (F) or
antiferromagnetic (AF).  $e_0$ is   the ground-state energy per  spin,
$e_m$ is the mean energy at  infinite temperature (here $e_m=0$).  $n$
is the  highest  known order  of the  HT   expansion of  the  specific
heat. The     three last columns  are   parameters  extracted from our
analysis:  low-temperature  limit and position    of the peak.   These
result are obtained from the highest order  available.  The error bars
reflects the dispersion of the  different Pad{\'e} approximants at the
highest  order.  Figures with a  star  ($^*$) are those  for which the
exact value is known (see text).}
\end{center}
\label{tab:I}
\end{table*}
\end{widetext}

\section{Systems with thermally activated heat capacity}
\label{sec:cvgap}

\subsection{One-dimensional Ising model}

\begin{equation}
	H=2\sum_i S^z_i\cdot S^z_{i+1}
\end{equation}

\begin{figure}
\begin{center}
\resizebox{8cm}{!}{\includegraphics{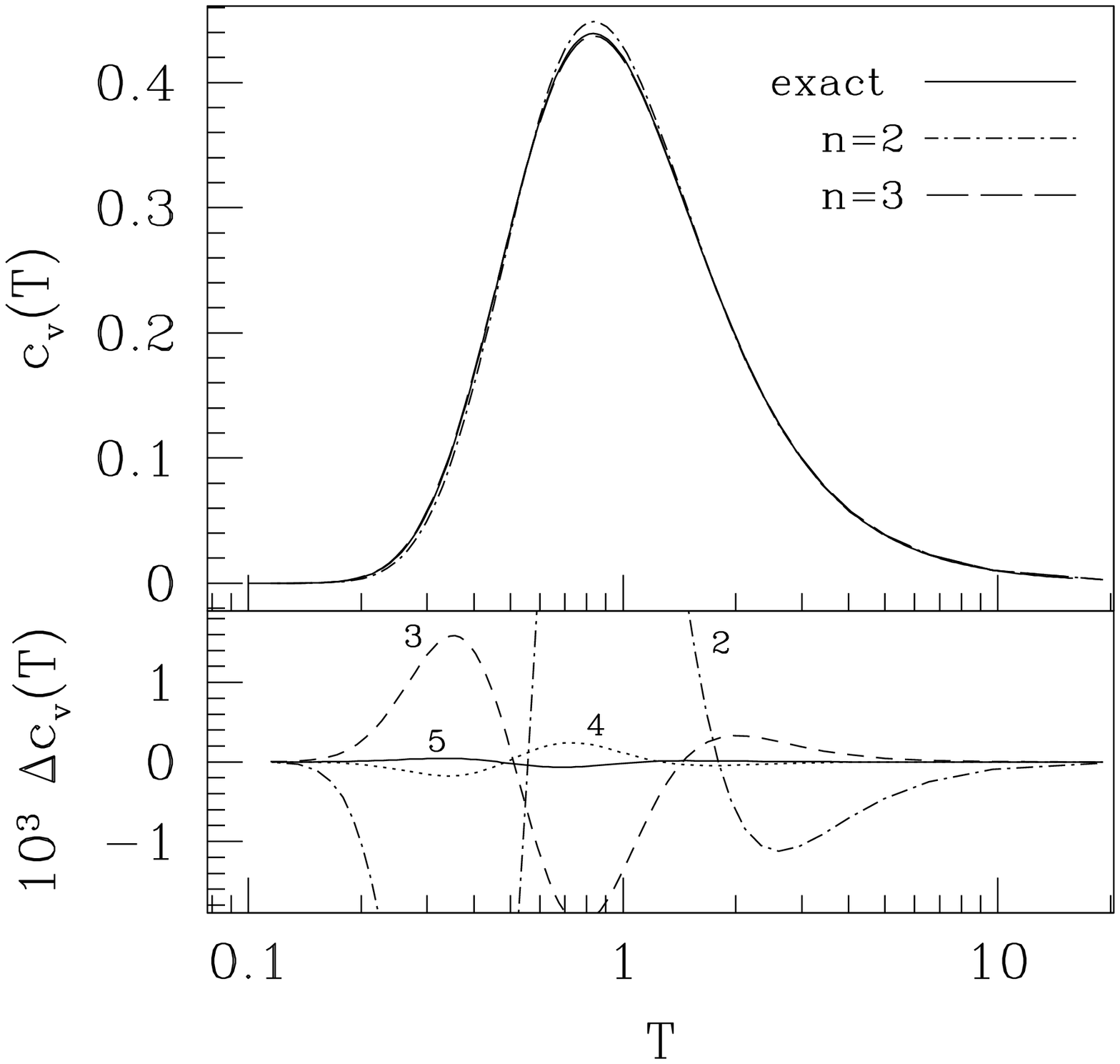}} 
\caption{{\bf One-dimensional Ising model}. 
{\em Top}: comparison between the exact specific  heat (full line) and
those obtained  from second (dot-dashed  line) and third (dashed line)
orders.   {\em  Bottom}:   difference  between  exact  result and  the
approximations at orders $n=2,3,4$ and 5.}
\label{fig:Ising}
\end{center}
\end{figure}

This model is the  simplest one with a gapped  spectrum. It  allows to
check the convergence of our method to the exact result in the case of
a low  temperature activated  heat capacity.   The  energy per site at
$T=0$  is $e_0=-1$ and $e_m=0$  at $T=\infty$.   The specific heat per
spin   is    given  by   $c_v(\beta=1/T)=(\beta/\cosh(\beta))^2$.  The
HT-expansion  starts   as $c_v(\beta=1/T)\sim \beta^2-\beta^4+\ldots$,
while   the spectrum  has  a gap  $\Delta=2$.    The heat capacity  is
thermally activated at low temperature:
\begin{equation}
	c_v(\beta\to\infty) \sim \beta^2\exp(-\Delta\beta)
\end{equation}
For such behavior, the entropy behaves as
\begin{equation}
	s(e)\sim-(e-e_0)\log(e-e_0)/\Delta
\label{eq:gapped_s}
\end{equation}
when $e\to  e_0$.   We  need  a  transformation   which converts  this
logarithmic  singularity into a    regular  behavior.  We choose   the
transformation 
\begin{equation}
	G(e)=(e-e_0)\frac{d}{d    e}\left(\frac{s(e)}{e-e_0}\right)
	\label{eq:Ggapped}
\end{equation}
and we approximate  $G$ by a Pad{\'e}  approximant (see appendix~C for
details).   The convergence is shown  in figure~\ref{fig:Ising} .  The
gap also converges rapidly  to the exact  value 2.  The  best Pad{\'e}
approximants are those of the form $[d,d]$ with $d\sim\frac{n}{2}$.

\subsection{One-dimensional $S=1$ Heisenberg model}

\begin{equation}
	H=\sum_i \vec{S}_i\cdot\vec{S}_{i+1}
\end{equation}
It is known that this  system exhibits a  spin gap (Haldane phase) and
the heat-capacity is thermally activated. More precisely, a non-linear
sigma model approach~\cite{jg94} gives:
\begin{equation}
	c_v(T)=\frac{\Delta}{\sqrt{2\pi}}
	\left(\frac{\Delta}{T}\right)^{\frac{3}{2}}
		\exp{\left(-\Delta/T\right)}
	\label{Cv_Haldane}
\end{equation} 
More generally,   if elementary excitations  are  weakly  interacting
massive bosons, one expects  $c_v(T)\sim \exp(-\Delta/T) T^{D/2-2}$ in
space dimension $D$.

\begin{figure}
\begin{center}
\resizebox{8cm}{!}{\includegraphics{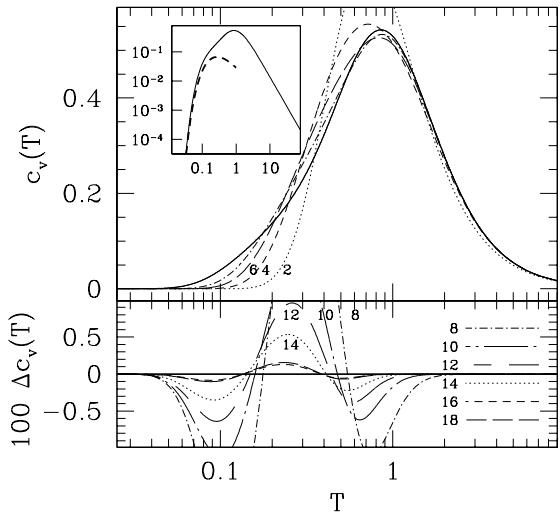}}
\caption{
Specific heat of the {\bf  antiferromagnetic Heisenberg spin-1 chain}.
{\em Top}: HT  orders are $n=2,4,8$  and 20 (full line).  {\em Inset}:
Log-log    plot   of   $n=20$     and  theoretical    prediction    of
Eq.~(\ref{Cv_Haldane}) (dashed).   {\em Bottom}:   Difference  between
$n=20$ and even $n$ from 8 to 18. $n=16,18$ differ from $n=20$ by less
than $2.10^{-3}$.}
\label{fig:Haldane}
\end{center}
\end{figure}

\begin{figure}
\begin{center}
\resizebox{8cm}{!}{\includegraphics{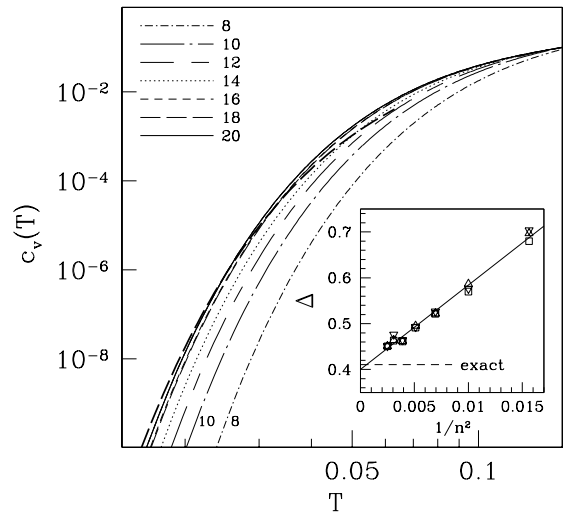}}
\caption{Convergence of $c_v(T)$ at low temperature
for the {\bf antiferromagnetic  Heisenberg spin-1 chain}.  Even orders
from 8 to 20 are shown. Full bold line is $n=20$, the dashed bold line
is from   Ref.~\cite{jg94}  (Eq.~\ref{Cv_Haldane}).  Notice that
$n=16$ and 18 are almost  identical at the scale of  the figure.  {\em
Inset}: Spin gap for  orders $n=8,10,12,14,16,18,20$.   The horizontal
dashed  line is   the exact value   (0.41) and   the  full  line is  a
least-square fit   to the data.    It extrapolates   to  $\Delta=.40$.
Upward   triangles      are   for    approximants      with    degrees
$[\frac{n}{2}+1,\frac{n}{2}-1]$         ,        square            for
$[\frac{n}{2},\frac{n}{2}]$     and  downward  triangles  for  degrees
$[\frac{n}{2}-1,\frac{n}{2}+1]$.  }
\label{fig:HaldaneGap}
\end{center}
\end{figure}

The  value of  the   ground-state energy and  spin  gap  is known very
accurately (from the density-matrix renormalization-group calculations
done by White and Huse~\cite{wh93}) to be $e_0=-1.401484038971(4)$ and
$\Delta=0.41050(2)$.  We  use the HT series data  obtained up to order
20 by Elstner, Jolic{\oe}ur  and Golinelli.~\cite{ejg94}  We apply the
same transformation  as   with   the  1D Ising  model.    Notice  that
Eq.~(\ref{eq:gapped_s})                       implies             $c_v(T)
\sim\exp{\left(-\Delta/T\right)}/T^2$ instead of $\sim
\exp{\left(-\Delta/T\right)}/T^{\frac{3}{2}}$ as     it should be from
Eq.~(\ref{Cv_Haldane}).  The  consequence of this little inconsistency
is that, at fixed order in the HT series, we overestimate the value of
the gap $\Delta$. Note that the correct low-T behavior would only give
a logarithmic correction to  $s(e)$ (Eq.~\ref{eq:gapped_s}).  This can
be rectified, in principle, by  a more sophisticated transformation of
$s(e)$.  Nevertheless, we obtain a gap which varies roughly as $1/n^2$
($n$ is the  order of  the HT-expansion)  and  the extrapolation shown
Fig.~\ref{fig:HaldaneGap} gives  $0.40$.  So, very  surprisingly, this
HT   approach  is able  to provide  some  quantitative and non-trivial
information about the low-energy physics.

\section{Conclusions}

We  have  presented a  new and   simple  method to  analyze HT  series
expansion for the specific heat  of spin systems.  It requires  only~:
the ground-state  energy and the qualitative  behavior of  $c_v(T)$ at
low  temperature.    These  two pieces  of information    allow  us to
constrain  the   specific   heat with  two   sum  rules  which improve
drastically the  convergence of the  HT expansion.  This  technique is
particularly  appropriate to analyze  specific  heat measurements in a
large temperature range. Since in many cases it is able to predict the
position and the  height of the maximum  of $c_v(T)$,  this method can
simplify and improve the determination of exchange parameter(s) and of
the number of  spins in the sample.   Remarkably, the method converges
down to zero temperature with only ten terms of  the HT series in most
of the cases we have investigated.   This method can even provide some
quantitative information on the low-energy  physics (e.g. of the value
of the spin gap of the  Haldane chain, or the zero-temperature entropy
of the Kagome antiferromagnet.~\cite{bmKag00}).

It would be interesting to  apply this technique to more sophisticated
models such as  the $t-J$ or  Hubbard models. It also seems worthwhile
to  investigate other interpolation  methods  than the simple Pad{\'e}
approximants presented here.  The  application of this new HT analysis
to  systems   with  finite-temperature  phase  transitions is  another
promising direction we are currently investigating.

{\bf Acknowledgments}  We are grateful N.~Elstner, T.~Jolic{\oe}ur and
O.~Golinelli for providing us with  their unpublished HT series on the
$S=1$ Heisenberg   chain.      It  is  also   a  pleasure    to  thank
T.~Jolic{\oe}ur and C.~Lhuillier, M.~Roger  and P.~Sindzingre for many
valuable  discussions  and J.~Talbot for    a careful reading  of  the
manuscript.


\appendix

\section*{Appendix A: Series for $s(e\to0)$}\label{A:series}

We  describe how the series  for $s(e\to0)$  is obtained from the
series for  $c_v(T\to\infty)$.   Assume the expansion of  the specific
heat is known up to order $n$:
\begin{equation}
	c_v(T)_{T\to\infty}
	=\sum_{i=2}^n{\frac{a_i}{T^i}} +\mathcal{O}(\frac{1}{T^{n+1}})
\end{equation} 
Eqs.~(\ref{eq:Defbeta}) and (\ref{eq:DefCv}) implies
\begin{eqnarray}
	s''(e)c_v\left(T=1/s'(e)\right)=-\left[s'(e)\right]^2 \nonumber \\
	s''(e)\left[
		\sum_{i=2}^n{ a_i \; s'(e)^i}
	\right] =-\left[s'(e)\right]^2 \label{eq:Devs}
\end{eqnarray}
Expanding  Eq.~(\ref{eq:Devs}) in  powers of  $e$  gives the series for
$s(e\to0)$.  For instance, if we have
\begin{equation}
	c_v(T)_{T\to\infty} 	= \frac{a_2}{T^2}
				+\frac{a_3}{T^3}+\frac{a_4}{T^4}
				+\mathcal{O}(\frac{1}{T^5})
\end{equation}
we obtain
\begin{equation}
	s(e)_{e\to0}=	\ln(2)
			-\frac{1}{2a_2}e^2
			-\frac{a_3}{6a_2^3}e^3
			+\frac{2a_4a_2-3a_3^2}{24a_2^5}e^4
			+\mathcal{O}(e^5)
	\label{eq:gapprox}
\end{equation}

In fact,  solving Eq.~(\ref{eq:Devs}) can be  done  very simply with a
software like {\sc Maple}:\\
	\hspace*{1cm}! n  := 4; Order:=n+1:  !\\
	\hspace*{1cm}! cv := !{\bf add}! ( a[i]/T**i, i=2..n );!\\
	\hspace*{1cm}! eq := -!{\bf D}!(s)(e)**2 / !{\bf D}!(!{\bf D}!(s))(e) ! \\
	\hspace*{3cm}		!= !{\bf subs}!(T=1/!{\bf D}!(s)(e),cv);!\\
	\hspace*{1cm} {\bf dsolve}!(!\{!eq, s(0)=ln(2), !{\bf D}!(s)(0) = 0!\} !,!\\
	\hspace*{3cm}		!s(e),'!{\bf type}!=!{\bf series}!' ) ;!\\

\section*{Appendix B: Pad{\'e} approximant for $G(e)$}\label{A:PadeG}

When the  system has a specific heat  with a power law $T^\frac{p}{q}$
at low  temperature the function  $G(e)=s(e)^{p+q}$ is regular  at the
ground-state energy $e=e_0$  as well as at  high temperature $e=0$. We
can therefore approximate $G$ by:
\begin{equation}
\label{eq:GeneralG}
	G^{\rm app}(e)=
	\s^{p+q}\left(1-\frac{e}{e_0}\right)^p {\rm Pade}_{[u,d]}(e)
	\label{eq:GappPade}
\end{equation}
where  ${\rm   Pade}_{[u,d]}(e)=N(e)/D(e)$,  $N$  (resp.  $D$)    is a
polynomial   of degree   $u$  (resp.   $d$)    and  $N(0)=D(0)=1$.  By
construction,  Eq.~(\ref{eq:GeneralG})  guarantees that  $G^{\rm app}$
has    the   correct     behavior   at    low   temperature    $G^{\rm
app}\sim(e-e_0)^p$. One has  to  evaluate the Pad{\'e} approximant  so
that the HT expansion  of $G(e)$ (and thus  of $s(e)$ and $c_v(T)$) is
exact up to order  $n=u+d$.  As usual, this is  done by expanding both
sides of Eq.~(\ref{eq:GappPade}) in powers of $e$ and solving a linear
system to determine the  unknown  coefficients of the  two polynomials
$N$ and $D$.  As an  example, we give  here the general expression for
the three possible approximants at order $n=2$:
\begin{equation}
	{\rm Pade}_{[2,0]}(e) =
	1+p\frac{e}{e_0}
	+\frac{1}{2} \left[ p(p+1)-x
			\right]
			\left(\frac{e}{e_0}\right)^2
\end{equation}

\begin{eqnarray}
	{\rm Pade}_{[1,1]}(e) =\frac{
		1+\frac{1}{2}\left[
			p-1+\frac{x}{p}
		\right]\frac{e}{e_0}
		}{
		1+\frac{1}{2}\left[
			-p-1+\frac{x}{p}
		\right]\frac{e}{e_0}
		}
\end{eqnarray}

\begin{equation}
	{\rm Pade}_{[0,2]}(e)^{-1} =
	1-p\frac{e}{e_0}
	+\frac{1}{2} \left[ p(p-1)+x
			\right]
			\left(\frac{e}{e_0}\right)^2
\end{equation}
where
\begin{equation}
	x=\frac{e_0^2}{a_2 }\;\frac{p+q}{\ln(2)}
\end{equation}

\section*{Appendix C: $G(e)$ for gapped systems}
\label{A:PadeGgap}
With gapped systems, the specific heat has the form:
$c_v(T)=\exp(\Delta/T)T^{\alpha-2}$.
We obtain the low energy limit:
\begin{equation}
	s(e)\simeq-\frac{e-e_0}{\Delta}\left[ \ln((e-e_0)\Delta)
	+\alpha\ln(\frac{-ln(e-e_0)}{\Delta})\right]
\end{equation}
For   $\alpha=0$,  we   recover    Eq.~(\ref{eq:gapped_s})  and    the
transformation of  Eq.~(\ref{eq:Ggapped}) holds.  When $\alpha\neq 0$,
in  order  to   remove  the logarithmic   singularities,  one   has to
differentiate  several    times.   But such   transformations  provide
singular behavior for $s(e)$, mainly because  of the change of sign in
the arguments of the log between $e=e_0$ and $e=e_m=0$.  More work has
to be done in these cases.

Unlike the case of a power-low behavior at low temperature, cases with
a gap require an integration in  order to go  back from $G(e)$ defined
Eq.~(\ref{eq:Ggapped}) to the entropy and the  specific heat. Since we
look for approximations of $G(e)$ in a Pad{\'e} form, this integration
can be  performed analytically.  The value of  the gap $\Delta$ can be
obtained   directly    from   $G$,     without   integration,    since
$G(e=e_0)=\frac{1}{\Delta}$.


\begin{thebibliography}{}

\bibitem{guttmann89}
A. J. Guttmann,  in  {\it Phase Transitions and   Critical Phenomena},
edited by C.~Domb and   J.~Lewovitz (Academic Press, New York,  1989),
Vol.~13 (1989).
\bibitem{note1}
For  the heat capacity, here  is   a non-exhaustive list of  available
series for quantum  models: Order   $\beta^{24}$ for  the   Heisenberg
spin-$\frac{1}{2}$     chain.~\cite{beu00} Order    $\beta^7$  for the
Heisenberg chain  with  arbitrary  value  of   the spin   $S$.   Order
$\beta^{10}$   for    a     frustrated   Heisenberg spin-$\frac{1}{2}$
chain.~\cite{beu00}   Order  $\beta^{20}$  for the  Heisenberg  spin-1
chain.~\cite{ejg94}   Order $\beta^{12}$   for  two-   and   three-leg
ladders.~\cite{osz96}  Order    $\beta^{14}$ for  square,  triangular,
simple  cubic, bcc and  fcc lattices.~\cite{Oitmaa} Order $\beta^{16}$
for  the Heisenberg  model  on  the Kagome lattice.~\cite{ey94}  Order
$\beta^7$  for   the   spin-$\frac{1}{2}$   Heisenberg  model  on  the
Shastry-Sutherland  lattice.~\cite{zho99}   Order $\beta^6$  for   the
$J_2-J_4-J_5$  multiple-spin   exchange    model on   the   triangular
lattice.~\cite{roger98}  Order $\beta^7$  for the  single-band Hubbard
model  on several  two-  and  three-dimensional lattices.~\cite{hoa92}
Numerous series for classical Ising-like  models are also available in
the literature.
\bibitem{roger98b}
M.~Roger.
Phys.  Rev.  B {\bf 58}, 11115 (1998).
\bibitem{narayanan90}
R.~Narayanan and R.~R.~P.~Singh.
Phys. Rev. {\bf B 42}, 10305 (1990).
\bibitem{zho99}
W.~Zheng,   C.~J.~Hamer and  J.~Oitmaa.
Phys. Rev. B {\bf 60}, 6608 (1999).
\bibitem{wang92}
J.~Wang.
Phys. Rev. B {\bf 45}, 2282 (1992).
\bibitem{Neef74}
T.~de~Neef.
Phys. Lett. {\bf 47}A, 51, (1974).
\bibitem{nj75}
T.~de~Neef   and  J.~M.~de Jonge.
Phys. Rev. B {\bf 11}, 4402 (1975).
\bibitem{jaklic96}
J.~Jaklic and P.~Prelovsek. 
Phys. Rev. Lett. {\bf 77}, 892 (1996).
\bibitem{BonnerFisher64}
J.~C.~Bonner and M.~E.~Fisher,
Phys. Rev. {\bf 135}A, 640 (1964).
\bibitem{nkk74}
T.~de~Neef, A.~J.~M.~Kuipers  and  K.~Kopinga.
J. Phys A. {\bf 7}, L171 (1974).
\bibitem{blote75}
H.~W.~Bl{\"o}te.
Physica {\bf 79}B, 427 (1975).
\bibitem{imada87}
M.~Imada.
J. Phys. Soc. Jpn {\bf 56}, 311 (1987).
\bibitem{makivic91}
M.~Makivic   and  H-Q.~Ding.
Phys.  Rev. B {\bf 43}, 3562 (1991).
\bibitem{note2}
In  particular, they prevent  the heat  capacity from  diverging in an
unphysical way at  low temperature, as  it is  often  the case in  the
standard Pad{\'e} approach where spurious poles can appear on the real
axis.
\bibitem{note3}
Let  $\rho_N(E)$ be the  density of states of  a large system with $N$
sites.  In  the micro-canonical ensemble,  when the energy per site is
$e=E/N$,   the          entropy         is      proportional        to
$s(e)=\frac{1}{N}\log{\rho_N(E=eN)}$.  A saddle-point expansion of the
partition  function $Z(\beta)=\int{\rho(E)e^{-\beta E}dE}$ when $N \to
\infty$  gives Eq.~(\ref{eq:Defbeta})    and  a free energy   density:
$F=-\frac{1}{N}T\log{Z}=e+Ts(e)$.
\bibitem{note4}
The method  indeed applies to many  other models, including those with
charge degrees of freedom: $t-J$, Hubbard model, etc...
\bibitem{note5}
In   principle, an exponential  ground-state  degeneracy  would give a
non-zero entropy at $T=0$.  It is not the case in  the models we study
here.  However, the Kagome-lattice antiferromagnet~\cite{ThermoKago00}
could be an  example   with a non-zero   ground-state  entropy.   Some
preliminary results  obtained with present  method indicate that it is
indeed the case.~\cite{bmKag00}.
\bibitem{k62}
S.~Katsura.
Phys. Rev. {\bf 127}, 1508 (1962).
\bibitem{bethe31}
H.~A.~Bethe, Z. Phys. {\bf 71}, 205 (1931).
\bibitem{h38}
L.~Hulth{\'e}n, Arkiv Mat. Astron. Fysik {\bf 26A}, 1 (1938).
\bibitem{a86}
I.~Affleck.
Phys. Rev. Lett. {\bf 56}, 746 (1986).
\bibitem{brg64}
G.~A.~Baker, G.~S.~Rushbrook   and   H.~E.~Gilbert.
Phys. Rev.  {\bf135A}, 1272 (1964).
\bibitem{beu00}
A.~B{\"u}hler, N.~Elstner,  G.~S.~Uhrig.
Eur. Phys. J. B {\bf 16}, 475 (2000).
\bibitem{takahashi}
M.~Takahashi,  Prog. Theor. Phys.   {\bf 46}, 401  (1971);
{\it ibid.} {\bf 50}, 1519 (1973).
\bibitem{klumper}
A.~Kl{\"u}mper.
Eur. Phys. J. B {\bf 5}, 677 (1998).  A.~Kl{\"u}mper and D.~C.~Johnston.
cond-mat/0002140 (2000).
\bibitem{ElstnerHT}
N.~Elstner, R.~R.~P.~Singh, A.~P.~Young.
Phys. Rev. Lett. {\bf 71} 1629 (1993).
\bibitem{ishida97}
K. Ishida, M. Morishita, K. Yawata, and H. Fukuyama.
Phys. Rev. Lett. {\bf 79}, 3451 (1997).
\bibitem{bllp94}
B.~Bernu, P.~Lecheminant, C.~Lhuillier, and L.~Pierre.
Phys.   Rev.  B {\bf  50}, 10048 (1994).
\bibitem{lblp95a}
P.~Lecheminant and  B.~Bernu  and C.~Lhuillier  and L.~Pierre.
Phys.  Rev.  B {\bf 52}, 9162 (1995).
\bibitem{takahashi89}
M.~Takahashi.
Phys. Rev. B {\bf 40}, 2494 (1989).
\bibitem{jg94}
T.~Jolic{\oe}ur and O.~Golinelli.
Phys. Rev. B {\bf 50}, 9265 (1994).
\bibitem{wh93}
S.~R.~White and  D.~A.~Huse.
Phys Rev. B {\bf 48}, 3844 (1993).
\bibitem{ejg94}
N.~Elstner, T.~Jolic{\oe}ur and O.~Golinelli. Unpublished (1994).
\bibitem{bmKag00}
B.~Bernu and   G.~Misguich.
To be published (2000).
\bibitem{osz96}
J. Oitmaa, R. R. P. Singh, and W. Zheng.
Phys. Rev. B {\bf 54}, 1009 (1996).
\bibitem{Oitmaa}
J.~Oitmaa, E.~Bornilla
Phys. Rev. B {\bf 21} 14228 (1996).
\bibitem{tc90}
N.~Trivedi and D.~M.~Ceperley.
Phys. Rev. B {\bf 41}, 4552 (1990).
\bibitem{bs98}
M.~C.~Buonaura and S.~Sorella.
Phys. Rev. B {\bf 57}, 11446 (1998).
\bibitem{ey94}
N.~Elstner    and  A.~P.~Young.
Phys.  Rev. B {\bf 50}, 6871 (1994).
\bibitem{roger98}
M.~Roger,  C.~B{\"a}uerle, Yu.~M.~Bunkov, A.-S.~Chen, and  H.~Godfrin.
Phys. Rev. Lett. {\bf 80}, 1308 (1998).
\bibitem{hoa92}
J. A. Henderson, J. Oitmaa, and M. C. B. Ashley. 
Phys. Rev. B {\bf 46}, 6328 (1992).
\bibitem{ThermoKago00}
P.~Sindzingre,     G.~Misguich, C.~Lhuillier,   B.~Bernu,   L.~Pierre,
C.~Waldtmann, H.-U.~Everts.
Phys. Rev. Lett., {\bf 84} 2953 (2000).

\end{thebibliography}
\end{document}